\documentclass[preprint,12pt]{elsarticle}

\usepackage{amssymb,amsmath,amsfonts,amsthm}
\usepackage{color}
\newcommand{\bracket}[1]{\left\langle #1 \right\rangle}

\journal{Physica A}

\begin{document}

\begin{frontmatter}



\title{Viral spreading of daily information in online social networks}


\author[1]{Tatsuro Kawamoto}
\author[2]{Naomichi Hatano}

\address[1]{Department of Physics, The University of Tokyo, Komaba, Meguro, Tokyo 153-8505, Japan}
\address[2]{Institute of Industrial Science, The University of Tokyo, Komaba, Meguro, Tokyo 153-8505, Japan}

\begin{abstract}
We explain a possible mechanism of an information spreading on a network which spreads extremely far from a seed node,
namely the viral spreading. 
On the basis of a model of the information spreading in an online social network, in which the dynamics is expressed as a random multiplicative process of the spreading rates, 
we will show that the correlation between the spreading rates enhances the chance of the viral spreading, 
shifting the tipping point at which the spreading goes viral. 
\end{abstract}


\end{frontmatter}

% \linenumbers


\section{Introduction}
A famous phenomenon on some types of online social networks such as Twitter and Facebook in which 
a post by a single user collects enormous attention, or `goes viral', may reflect 
structural and dynamical properties which we have not seen in other conventional networks on the web.
Not only that the information flow on the web became very active in the last decade, 
but web services which people use to broadcast and receive the information have also changed greatly. 
It is important to investigate what their characteristic properties are and how they affect the information flow in order to predict their behaviors.
Our aim here is to understand the mechanism of typical information spreading in an online social network with the most simple modeling. 

There are various types of information spreading in the web and many ways to observe such phenomena; 
we can consider models with many levels of precision \cite{Newman02,Liben-Nowell08,Golub10,Iribarren09,Iribarren11,Vazquez06,Vazquez07,WuHuberman07,Wilkinson08,Yan11,Galuba10,Bakshy11,Wu11,kwmt}. 
One of the conventional ways in which the spreading of information occurs in the web is due to 
%The conventional information spreading on the web occurs owing to 
the access of users to the spreaders of the information, \textit{e.g.} web-news, Wikipedia, blogs, \textit{etc}. \cite{WuHuberman07,Wilkinson08,Yan11}. 
On the other hand, a different kind of information spreading in the web is getting increasingly commonplace; 
the major examples are \textit{retweet} of Twitter and \textit{share} on Facebook. 
The information spreading in such online social networks is qualitatively different from the former one; 
instead of accessing to the spreaders, users receive the information passively and transmit it to other users, 
thereby helping the information to diffuse. 

In the previous work \cite{kwmt}, one of the authors constructed a local spreading model to describe the typical behavior of such an information spreading process. 
In the present paper, we will focus on the situation where the spreading goes viral, \textit{i.e.}, the information which spreads to users who are extremely far from a seed user. 
As we did in the previous work, we will consider the case of the tweet spreading on Twitter as an example. 
The situation that we imagine for a viral spreading is the spreading of information of general interest, \textit{e.g.}, postings with funny jokes, poetic writings, important news which are not broadcasted on other mass media, \textit{etc}. 
The higher the fraction of the retweeters among the viewers of the tweet (we call it \textit{spreading rate} or \textit{retweet rate}) is, the wider the range of the spreading is, which also results in a large number of retweets.

A naive description of a tweet which enjoys many retweets would be the retweeting by a single user with a large number of the followers. 
Although it might be an important factor, even the accounts with millions of followers do not receive thousands of retweets for their daily tweets. 
Therefore, such a naive description does not explain the whole mechanism of the viral spreading. 
The cooperation by many users is presumably crucial to the spreading of the tweet.

Let us define a viral spreading more precisely. 
%In this paper, 
We assume a tree structure with a homogeneous degree distribution for the underlying network and an infinite path length from a seed user. 
Mathematically, we define a viral spreading as the spreading which never stops in such a network; 
there exists a tipping point for the spreading rate at which the spreading goes viral. 
In spite of the highly clustered structure of the online social network, the validity of the tree approximation for a spreading path has been theoretically considered and empirically proven in Refs.~\cite{Newman02,Iribarren11,Vazquez07} and the references therein. 
Although the network is full of loops, it does not necessarily mean that the spreading path contains many loops.
Our goal here is not to reproduce the statistical behavior of the data precisely, 
but to explore mathematical properties of the semi-microscopic law of spreading; 
even though spreadings always die out in reality because of the finite path length as well as the decay of the spreading rate due to the temporal effect and the distance from the seed user, 
the analysis of such a tipping point on the present toy model seems a plausible guideline for a spreading to go viral.

As we will mention later, we will treat the spreading rate in our spreading model as a stochastic variable which obeys a lognormal distribution. 
In the previous work \cite{kwmt}, we neglected the effect of correlation between the spreading rates of the followers. 
Whenever the spreading goes viral, however, we can easily imagine that the effect of (positive) correlation plays an important role. 
That is, if some nodes contribute to the spreading, the receivers of the information from those nodes tend to contribute to the spreading as well. 
We will treat such an effect as a perturbation from the independent process. 
Because the stochastic variables obey lognormal distributions which are fat-tail distributions, 
it is interesting to see how their correlation affects the dynamics.
We will show that indeed it can largely enhance the chance of the viral spreading.

This paper is organized as follows. 
After describing the spreading model which we consider, 
we discuss the tipping point of the viral spreading in the case of independent spreading rates. 
Then we show that the tipping point is shifted owing to the correlation between the spreading rates.

\begin{figure}[t]
\begin{center}
\includegraphics[width=0.6 \textwidth]{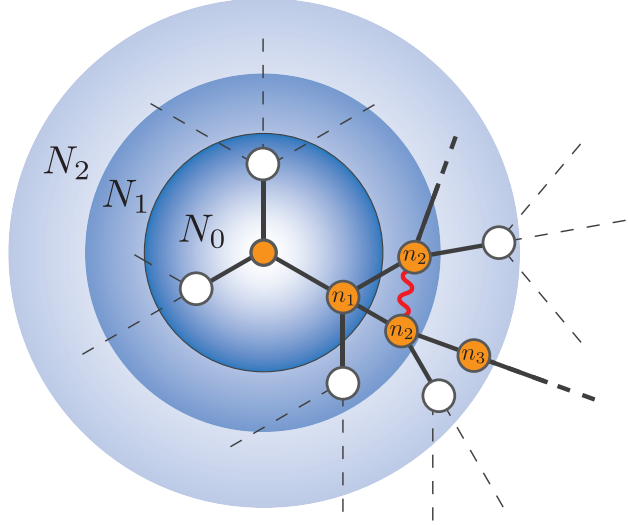}
\end{center}
\caption{
(Color online) 
Information spreading on an online social network. 
The node at the center represents the seed and the linked nodes can receive the information.
A solid line represents that the information has diffused through the link.
We ignore the over-counting of nodes such as the one illustrated by the wavy line; 
\textit{i.e.}, we assume a tree structure. 
}
\label{TwitterNetwork}
\end{figure}

\section{Model}
In order to model the information spreading in an online social network, 
we classify the informed nodes by the distance from the seed node; 
see Fig.~\ref{TwitterNetwork}.
We call the nodes with the same distance a \textit{generation} and discuss the spreading process with respect to the generations. 
We denote the number of nodes in the $g$th generation by $N_{g}$. 
Among the $N_{0}$ nodes which are directly connected to the seed node, some of them contribute to the spreading and pass the information to the nodes in the first generation. 
Because we assumed that the base graph is a loopless tree with a homogeneous degree distribution, 
we estimate the number $N_{1}$ of the nodes in the first generation as 
\begin{align}
&N_{1} = \overline{k} \beta_{1} N_{0} =: J_{1} N_{0}, 
\end{align}
where $\beta_{1}$ is the spreading rate which is a positive stochastic variable indicating the rate of nodes contributing to the spreading among the $N_{0}$ nodes and $\overline{k}$ is the average number of links to a node; in the case of Twitter, the spreading rate $\beta_{1}$ is the retweet rate and $\overline{k}$ is the average number of followers. 
Applying this process to all generations, we obtain the following random multiplicative process: 
\begin{align}
&N_{m} = N_{0} \prod_{g=1}^{m} J_{g} = N_{0} \overline{k}^{m} \prod_{g=1}^{m} \beta_{g} 
\end{align}
for $m \ge 1$.
Summing up $N_{m}$ for all $m$, we obtain the total number of viewers $N_{\mathrm{tot}}$ as 
\begin{align}
%N_{\mathrm{tot}} = N_{0} \left\{ 1 + J_{1}\left[ 1 + J_{2} (1 + \cdots ) \right] \right\}, \label{Ntot}
N_{\mathrm{tot}} = N_{0} \left( 1 + \sum_{m=1}^{\infty} \prod_{g=1}^{m} J_{g} \right). \label{Ntot}
\end{align}
The tipping point of the viral spreading is the point where $N_{\mathrm{tot}}$ diverges.

For simplicity, we assume that every spreading rate $\beta_{g}$ obeys a common probability distribution.
Especially in the case of Twitter, 
we confirmed \cite{kwmt} that $\beta_{g}$ roughly obeys a common lognormal distribution for $g=2$ irrespective of the seed node, 
while its average and variance depend on the character of the seed node for $g=1$. 
Then we set the distribution of the stochastic variable $J_{g} = \overline{k}\beta_{g}$ to be 
\begin{align}
p(J_{g}) = \frac{1}{J_{g}\sqrt{2\pi \sigma^{2}}} \exp \left[ -\frac{1}{2\sigma^{2}} (\ln J_{g} - \mu)^{2} \right], \label{logN}
\end{align}
and express $J_{g}$ as 
\begin{align}
J_{g} = \mathrm{e}^{\mu + \xi_{g}}, \label{J-xi}
\end{align}
where $\mu$ and $\sigma^{2}$ are constant and $\xi$ is a stochastic variable which obeys a Gaussian distribution $\mathcal{N}(0,\sigma^{2})$.

The crucial point of our model is that the stochastic variable $\beta_{g}$ is assigned to each generation. 
In the case of the branching process \cite{Iribarren09,Iribarren11,Vazquez06,Vazquez07,HarrisBook,AthreyaBook}, a stochastic variable is assigned to each node; 
it determines whether a node contributes to the spreading (and how many descendants it produces). 
The histogram of the spreading rate would be a binomial distribution in the branching process, as long as each stochastic variable obeys an identical distribution, because each node in a generation contributes to the spreading with a certain probability $q$ and does not with probability $1-q$. 
Hence, we gave up assigning a stochastic variable to each node and instead considered a stochastic variable for each generation so that the spreading rate may obey an arbitrary probability distribution.

The idea of dividing the nodes in the network into the generations as above itself already exists in the literature 
\cite{Newman2003,FontouraCosta2006,CostaRocha2006,Baronchelli2006}. 
It is often called a \textit{ring}. 
For example, Baronchelli \textit{et al.} \cite{Baronchelli2006} analyzed the mean first passage time of a random walker using the ring structure. 
On our model, however, there is no cycle flow and the spreading rate
%, i.e. the retweet rate, 
is not a constant but obeys a fat-tail distribution. 
Hence, the behavior of our model should be different from a simple random walk or a fundamental branching process.

\section{The case of independent spreading rate}
In the following, we will consider the average number of the informed nodes $N_{\mathrm{tot}}$, normalized by $N_{0}$. 
In the case where the stochastic variables $J_{g}$ are independent of each other and all their averages are the same, \textit{i.e.} $\bracket{J_{g}} = \bracket{J}$, we have 
\begin{align}
\frac{ \bracket{N_{\mathrm{tot}}} }{ N_{0} } 
&= 1 + \bracket{J} + \bracket{J}^{2} + \bracket{J}^{3} + \cdots \nonumber\\
&= \frac{ 1 }{ 1-\bracket{J} } \label{Independent1}
\end{align}
for $\bracket{J} < 1$, 
where $\bracket{\cdots}$ stands for the statistical average with respect to the distribution of the stochastic variables. 
In the case of the lognormal distribution (\ref{logN}), 
we have $\bracket{J} = \exp\left(\mu +  \sigma^{2}/2 \right)$.
%the average $\bracket{J}$ reads 
%\begin{align}
%\bracket{J} 
%&= \mathrm{e}^{\mu} \exp \left[ \ln \bracket{ \mathrm{e}^{ \xi} } \right]
%= \mathrm{e}^{\mu} \exp \left[ \bracket{ \xi }_{\mathrm{c}} + \frac{1}{2} \bracket{ \xi^{2} }_{\mathrm{c}} \right] \nonumber\\
%&= \exp\left(\mu + \frac{ \sigma^{2} }{ 2 } \right), 
%\end{align}
%where $\bracket{\cdots}_{\mathrm{c}}$ is the cumulant. 

Since $J_{g} = \beta_{g} \overline{k}$, and hence $\bracket{J} = \bracket{\beta} \overline{k}$, 
Eq.~(\ref{Independent1}) gives the tipping point 
\begin{align}
\beta_{\mathrm{ex}} = \overline{k}^{-1} \label{Independent2}
\end{align}
for the viral spreading. 
In the case of the Twitter network, $\overline{k} \sim \mathcal{O}(10^{2})$
and hence the tipping point is $\beta_{\mathrm{ex}} \sim \mathcal{O}(10^{-2})$. 
On the other hand, in the case of some major news accounts such as The New York Times (@nytimes) and Reuters Top News (@Reuters), 
$\bracket{\beta} \sim \mathcal{O}(10^{-5})$, which is much lower than the tipping point. 
Because of the restriction of Twitter API \cite{API}, we cannot measure the value of the spreading rate $\beta_{g}$ of the viral spreading explicitly. 
Although the possibility of reaching the tipping point $\beta_{\mathrm{ex}} = \overline{k}^{-1}$ depends on the average and the variance of the spreading rate, the threshold appears to be too high to reach in reality if we assume that $J_{g}$ are independent of each other.

\section{The case of correlated spreading rates}
In order to make a better estimate of the tipping point, 
let us now consider the quantity $\bracket{N_{\mathrm{tot}}}/N_{0}$ in the case where the stochastic variables $J_{g}$ are not independent of each other. 
Instead of setting $\xi_{g}$ in Eq.~(\ref{J-xi}) as an independent Gaussian variable, we now set 
\begin{align}
&p(\{ \xi_{g} \}) = \frac{1}{Z} \exp\left[ -\frac{1}{2} \sum_{i j} \xi_{i} \Sigma^{-1}_{ij} \xi_{j} \right], 
&Z = \sqrt{\frac{(2\pi)^{N}}{\det \Sigma^{-1}}} , 
\label{Correlated1}
\end{align}
where $Z$ is the normalization factor and $\Sigma^{-1}$ is the inverse matrix of the covariance matrix $\Sigma_{ij} = \bracket{ \xi_{i}\xi_{j} }$. 
The matrix $\Sigma^{-1}$ is an infinite-dimensional matrix; 
we first treat it as an $N \times N$ matrix and take the limit $N\to \infty$ in the end. 
We assume the following matrix for $\Sigma^{-1}$: 
\begin{align}
\Sigma^{-1} &= 
\begin{bmatrix}
\sigma^{-2} &-\eta &0 &\cdots \\
-\eta &\sigma^{-2} &-\eta &\cdots \\
0 &-\eta &\sigma^{-2} &\cdots \\
\vdots &\vdots &\vdots &\ddots
\end{bmatrix}.
\end{align}

The statistical average of the number of the informed nodes is now given by 
\begin{align}
\frac{ \bracket{N_{\mathrm{tot}}} }{N_{0}}
&= 1 + \sum_{m=1}^{\infty} \bracket{ \prod_{g=1}^{m} J_{g}}, %\nonumber\\
%&= 1 + \sum_{m=1}^{\infty} \mathrm{e}^{m \mu} \bracket{ \exp \left( \sum_{g=1}^{m} \xi_{g} \right)}, \\
\end{align}
where the average $\bracket{\cdots}$ is now taken with respect to the correlated distribution (\ref{Correlated1}). 
In order to calculate the average, we diagonalize the matrix $\Sigma^{-1}$ with a unitary matrix $U$ to obtain 
\begin{align}
&P(\vec{x}) = \frac{1}{Z} \exp\left[ -\frac{1}{2} \sum_{i=1}^{N} (\sigma^{-2} - \eta \lambda_{i}) x^{2}_{i} \right], 
%& \Xi = \sqrt{\frac{(2\pi)^{N}}{ \prod_{i=1}^{N} (\sigma^{-2} - \eta \lambda_{i}) }} , 
\end{align}
where 
\begin{align}
&\vec{x} = U \vec{\xi}, \hspace{20pt} U_{mn} = \frac{1}{L} \sin(m k_{n}),\\
&\lambda_{\alpha} = 2 \cos k_{\alpha}, \hspace{20pt} k_{\alpha} = \frac{\pi \alpha}{N+1}, \\
& L^{2} = \frac{1}{2}(N + 1) .
\end{align}
After this diagonalization, we have 
\begin{align}
&\bracket{ \prod_{g=1}^{m} J_{g}}
%&\bracket{ \exp \left( \sum_{g=1}^{m} \xi_{g} \right)}
= \mathrm{e}^{m \mu} \int d\vec{\xi} \, P(\vec{\xi}) \exp \left( \sum_{g=1}^{m} \xi_{g} \right) \nonumber\\
&\hspace{10pt} = \mathrm{e}^{m \mu} \int \frac{d^{N}x}{Z} \, \exp\left[ -\frac{1}{2} \sum_{i=1}^{N} a_{i} x^{2}_{i} \right] 
\exp \left( \sum_{j=1}^{N} b_{j} x_{j} \right) \nonumber\\
&\hspace{10pt} = \mathrm{e}^{m \mu} \exp \left( \sum_{j=1}^{N} \frac{b^{2}_{j}}{2a_{j}} \right), \label{Jave}
\end{align}
where 
\begin{align}
a_{i} &= \sigma^{-2} - \eta \lambda_{i} = \sigma^{-2} - 2 \eta \cos k_{i}, \nonumber\\
b_{j} &= \frac{1}{L} \sum_{g=1}^{m} \sin(g k_{j}), 
\end{align}
and we used the relation 
\begin{align}
\sum_{g=1}^{m} \xi_{g} &= \frac{1}{L} \sum_{g=1}^{m} \sum_{j=1}^{N} \sin(g k_{j}) x_{j} = \sum_{j=1}^{N} b_{j}x_{j}. 
\end{align}
Substituting these values into Eq.~(\ref{Jave}), we obtain 
\begin{align}
&\bracket{ \prod_{g=1}^{m} J_{g}}
= \mathrm{e}^{m \mu} \exp \left[ \sum_{j=1}^{N} \sum_{g,g^{\prime}=1}^{m} \frac{ \sin gk_{j} \sin g^{\prime}k_{j} }{ a_{j}(N+1) } \right]\nonumber\\
&= \mathrm{e}^{m \mu} \exp \biggl[ \frac{1}{2(N+1)} \sum_{j=1}^{N} \sum_{g,g^{\prime}=1}^{m} 
a_{j}^{-1} \nonumber\\
&\hspace{50pt} \biggl( \cos k_{j}( g - g^{\prime} ) - \cos k_{j}( g + g^{\prime} ) \biggr) \biggr] \label{1stOrder-0}
\end{align}

Let us now consider the case where $\epsilon \equiv \eta / \sigma^{-2} \ll 1$ 
and analyze the expansion of $a_{j}^{-1}$ with respect to $\epsilon$: 
\begin{align}
a_{j}^{-1} %&= \left( \sigma^{-2} + 2 \eta \cos k_{j} \right)^{-1} \nonumber\\
%&= \sigma^{2} \left( 1- 2\epsilon \cos k_{j} \right)^{-1} \nonumber\\
&= \sigma^{2} \left( 1 + 2\epsilon \cos k_{j} + o(\epsilon) \right). \label{epsilon-expansion}
\end{align}
From the zeroth-order expansion, 
we simply obtain 
$\bracket{ \prod_{g=1}^{m} J_{g}} = \bracket{J}^{m}$, 
which reduces to the non-correlated case (\ref{Independent1}).
Including the first-order correction of $\epsilon$, we have 
\begin{align}
\bracket{ \prod_{g=1}^{m} J_{g}}
&= \bracket{J}^{m} \exp \left[ \frac{2\epsilon \sigma^{2}}{2(N+1)} \sum_{j=1}^{N} \sum_{g, g^{\prime} = 1}^{m}
\cos k_{j} \left( \cos k_{j}(g - g^{\prime}) - \cos k_{j}(g + g^{\prime}) \right) \right]. \label{1stOrder-1} %\nonumber\\
%&= \bracket{J}^{m} \exp \biggl[ \frac{\epsilon \sigma^{2}}{2(N+1)} \sum_{j=1}^{N} \sum_{g, g^{\prime} = 1}^{m}
%\bigl( \cos k_{j}(g - g^{\prime}+1) + \cos k_{j}(g - g^{\prime}-1)  \nonumber\\ 
%&\hspace{100pt} - \cos k_{j}(g + g^{\prime}+1) - \cos k_{j}(g + g^{\prime}-1) \bigr) \biggr]. \label{1stOrder-1}
\end{align}
%Noting that 
%\begin{align}
%&\sum_{j=1}^{N} \cos k_{j} \gamma 
%=\sum_{j=1}^{N} \cos \frac{j \pi}{N+1} \gamma \nonumber\\
%&= \begin{cases}
%0 &(\gamma = \text{odd}) \\
%-1 &(\gamma = \text{even}), 
%\end{cases}
%\end{align}
%then 
%\begin{align}
%&\sum_{j=1}^{N}\sum_{g, g^{\prime} = 1}^{m} \cos k_{j}(g + g^{\prime}+1)
%= \sum_{j=1}^{N}\sum_{g, g^{\prime} = 1}^{m} \cos k_{j}(g + g^{\prime}-1) \nonumber\\
%&= \begin{cases}
%-\frac{m^{2}}{2} &(m = \text{even}) \\
%-\frac{(m+1)(m-1)}{2} &(m = \text{odd}).
%\end{cases}
%\end{align}
%Also we have 
%\begin{align}
%&\sum_{j=1}^{N}\sum_{g, g^{\prime} = 1}^{m} \cos k_{j}(g - g^{\prime}+1) 
%= \sum_{j=1}^{N}\sum_{g, g^{\prime} = 1}^{m} \cos k_{j}(g - g^{\prime}-1) \nonumber\\
%&= 
%\begin{cases}
%N(m-1) -\left[ \frac{m^{2}}{2} - (m-1) \right] &(m=\text{even}) \nonumber\\ 
%N(m-1) -\left[ \frac{(m+1)(m-1)}{2} - (m-1) \right] &(m=\text{odd}). 
%\end{cases}
%\end{align}
%Therefore, Eq.~(\ref{1stOrder-1}) reads 
After some algebra, we obtain 
\begin{align}
\bracket{ \prod_{g=1}^{m} J_{g}}
%&= \bracket{J}^{m} \exp \biggl[ \frac{2 \epsilon \sigma^{2}}{2(N+1)} (N+1)(m-1) \biggl] \nonumber\\
&= \bracket{J}^{m} \mathrm{e}^{ \epsilon \sigma^{2}(m-1)}. \label{1stOrder-2}
\end{align}
Hence, the total number of the informed nodes normalized by $N_{0}$ reads 
\begin{align}
\frac{ \bracket{N_{\mathrm{tot}}} }{N_{0}}
&= 1 + \sum_{m=1}^{\infty} \bracket{ \prod_{g=1}^{m} J_{g}}
= 1 + \frac{ \bracket{J} }{ 1 - \bracket{J} \mathrm{e}^{ \epsilon \sigma^{2}} } \label{1stOrder-3}
\end{align}
for $\bracket{J} \mathrm{e}^{\epsilon \sigma^{2}} < 1$. 
Since $\bracket{J} = \bracket{\beta} \overline{k}$ again, the tipping point for the viral spreading $\beta_{\mathrm{ex}}$ now reads 
\begin{align}
\beta_{\mathrm{ex}}  = \overline{k}^{-1} \mathrm{e}^{-\epsilon \sigma^{2}} \label{threshold-epsilon}
\end{align}
instead of Eq.~(\ref{Independent2}). 
The correlation between the spreading rates thus shift the tipping point to a lower spreading rate. 

\begin{figure}[t]
\begin{minipage}{0.5\hsize}
\begin{center}
\includegraphics[width=\textwidth]{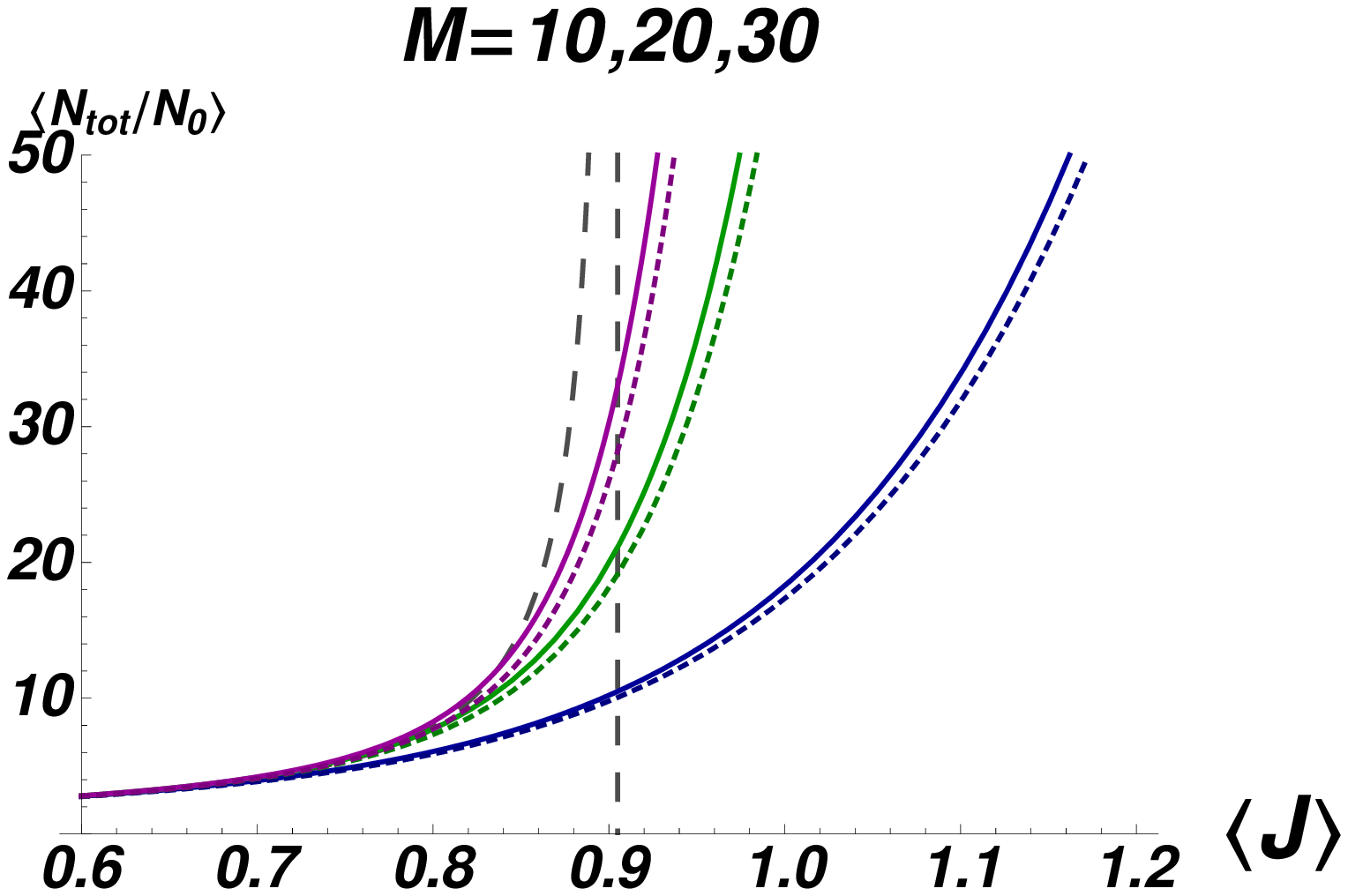}
(a)
\end{center}
\end{minipage}
\begin{minipage}{0.5\hsize}
\begin{center}
\includegraphics[width=\textwidth]{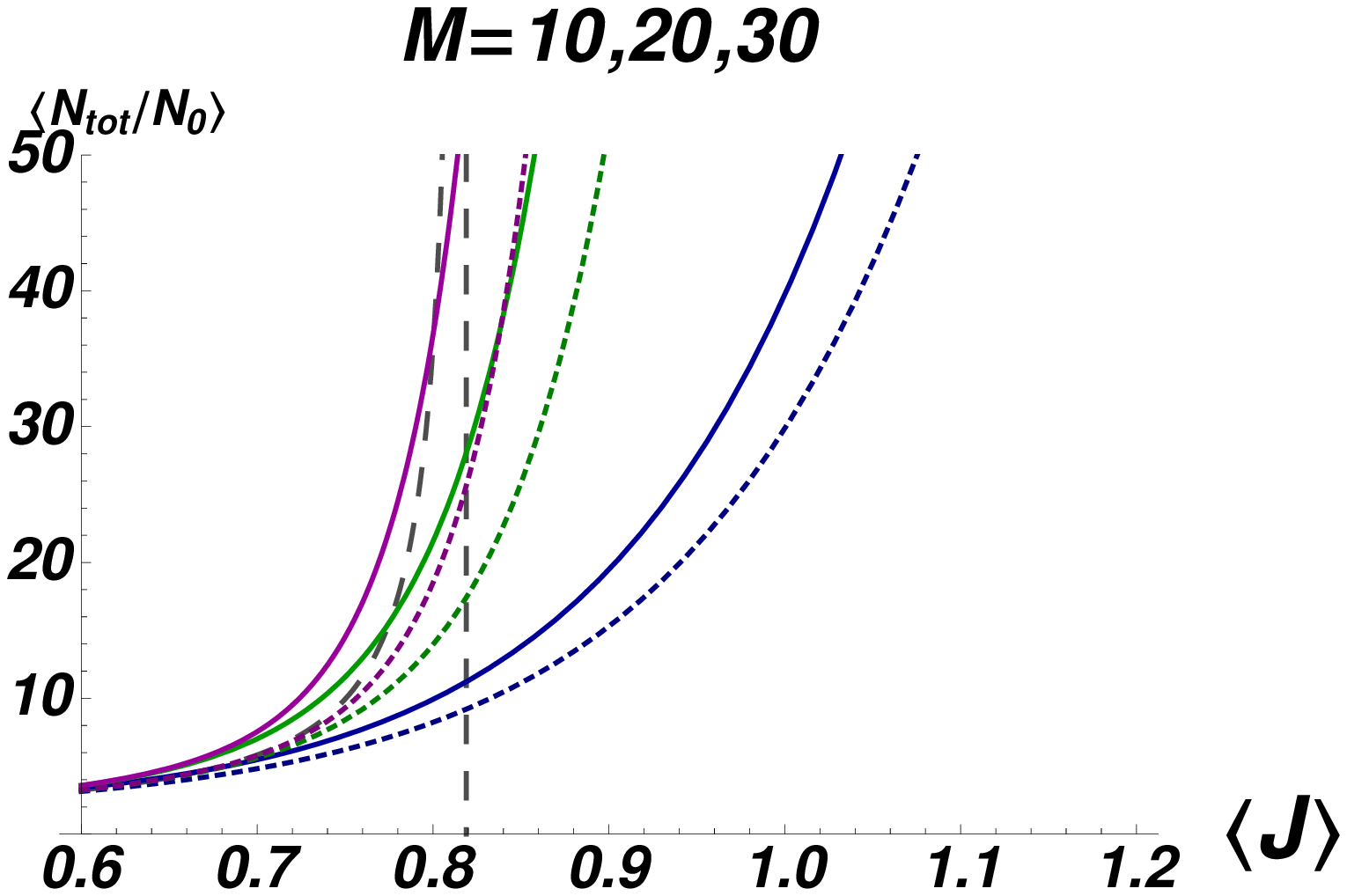}
(b)
\end{center}
\end{minipage}

\caption{
(Color online) 
Numerically calculated results of $\bracket{N_{\mathrm{tot}}}/N_{0}$ in Eq.~(\ref{M-sum}), where the sum is taken up to $M=10$ (right, blue), $20$ (middle, green), $30$ (left, red). 
The dotted lines indicate the approximated results with the perturbative estimate (\ref{1stOrder-2}) and the solid lines indicate the results with numerical estimates of Eq.~(\ref{1stOrder-0}). 
The parameters are set to $\sigma^{2} = 2$, $\epsilon = 0.05$ for (a) and $\sigma^{2} = 2$, $\epsilon = 0.1$ for (b). 
We set $N=30$ for the calculation of Eq.~(\ref{1stOrder-0}); the result is the same as long as $N \ge M$. 
The broken line shows the behavior of Eq.~(\ref{1stOrder-3}), which is the case of $M=\infty$ with the perturbative estimate (\ref{1stOrder-2}). 
}
\label{XplosionPlot}
\end{figure}

We expect that the perturbative estimate (\ref{1stOrder-2}) of the tipping point gives an upper bound of the true tipping point. 
In Fig.~\ref{XplosionPlot}, we can confirm it by comparing (i) (solid lines) numerical estimates of Eq.~(\ref{1stOrder-0}) substituted into 
\begin{align}
\frac{ \bracket{N_{\mathrm{tot}}}_{M} }{N_{0}}
&= 1 + \sum_{m=1}^{M} \bracket{ \prod_{g=1}^{m} J_{g}}, \label{M-sum}
\end{align}
and (ii) (dotted lines) perturbative estimates Eq.~(\ref{1stOrder-2}) substituted into Eq.~(\ref{M-sum}). 
The former is always greater than the latter as far as we checked. 
We hence expect that it is also true in the limit $M \rightarrow \infty$. 
Then the true curve of $\bracket{N_{\mathrm{tot}}}/N_{0}$ in the limit $M \rightarrow \infty$ should be greater than its perturbative estimate (\ref{1stOrder-3}) (dashed lines in Fig.~\ref{XplosionPlot}). 
It implies that the true tipping point of the viral spreading is equal to or lower than the perturbative estimate (\ref{threshold-epsilon}).

Let us next write down the tipping point in terms of the correlation coefficient of the spreading rates instead of the off-diagonal element $\epsilon = \eta / \sigma^{-2}$ of the matrix $\Sigma^{-1}$.
The matrix $\Sigma^{-1}$ which contains the off-diagonal element $\epsilon$ is the inverse matrix of the covariance matrix $\Sigma$ of $\xi_{g}$, 
which is related to that of $J_{g}$ by Eq.~(\ref{J-xi}). 
Expressing the inverse of the covariance matrix as $\Sigma^{-1} = \sigma^{-2} F_{N}$, the covariance matrix $\Sigma$ reads 
\begin{align}
&\Sigma_{ik} = \Sigma_{ki} = \frac{\sigma^{2}}{\det F_{N}} \det F_{i-1} \det F_{N-k} \, \epsilon^{k-i} \label{CovMat}
\end{align}
for $i \le k$, 
%\begin{align}
%\Sigma = \frac{\sigma^{2}}{\det F_{N}} 
%\begin{bmatrix}
%\det F_{N-1} &\epsilon \det F_{N-2} &\epsilon^{2} \det F_{N-3} &\cdots \\
%\epsilon \det F_{N-2} &\det F_{1}\det F_{N-2} &\epsilon \det F_{N-3} &\cdots \\
%\epsilon^{2} \det F_{N-3} &\epsilon \det F_{N-3} &\det F_{2}\det F_{N-3} &\cdots \\
%\vdots &\vdots &\vdots &\ddots \label{CovMat}
%\end{bmatrix}, 
%\end{align}
where the subscript of the matrix $F_{N}$ denotes the number of dimensions and we defined $\det F_{0} = 1$. 
The determinant of $F_{g}$ has the following recursion relation  
\begin{align}
\det F_{g} = \det F_{g-1} - \epsilon^{2} \det F_{g-2}. 
\end{align}
In the limit where $N \to \infty$, it reduces to 
%which reduces to 
\begin{align}
&\frac{1}{r} = 1 - \epsilon^{2} r, 
\end{align}
where 
\begin{align}
r= \lim_{N \rightarrow \infty} \frac{ \det F_{N-1} }{ \det F_{N} }. 
\end{align}
Considering the fact that $r$ needs to satisfy $r^{n} < \infty (n \to \infty)$,  we have
\begin{align}
&r = \frac{1 - \sqrt{1 - 4\epsilon^{2}} }{2\epsilon^{2}}. 
\end{align}
%and 
%\begin{align}
%\Sigma = \sigma^{2} 
%\begin{bmatrix}
%r &\epsilon r^{2} &\epsilon^{2} r^{3} &\cdots \\
%\epsilon r^{2} &r^{2} &\epsilon r^{2} &\cdots \\
%\epsilon^{2}r^{3} &\epsilon r^{2} &(1-\epsilon^{2})r^{3} &\cdots \\
%\vdots &\vdots &\vdots &\ddots
%\end{bmatrix}.
%\end{align}
Hereafter, we will work in the limit $N \rightarrow \infty$. 
Noting that $\det F_{g-1}$ and $r$ are both $1 + \mathcal{O}(\epsilon^{2})$, 
the matrix elements of Eq.~(\ref{CovMat}) read 
\begin{align}
%\bracket{ \xi_{g}^{2} } &= \sigma^{2} \det F_{g-1} \frac{\det F_{N-g}}{\det F_{N}} = \sigma^{2} \det F_{g-1} r^{g}, \\
%\bracket{ \xi_{g}\xi_{g+1} } &= \bracket{ \xi_{g+1}^{2} } \frac{\det F_{g-1}}{\det F_{g}} \, \epsilon. 
\bracket{ \xi_{g}\xi_{g+1} } &= \epsilon \sigma^{2} \det F_{g-1} \frac{\det F_{N-g-1}}{\det F_{N}} 
= \epsilon \sigma^{2} \det F_{g-1} r^{g+1} \nonumber\\
&= \epsilon \sigma^{2} + \mathcal{O}(\epsilon^{2}). 
\end{align}
Hence, up to the accuracy of $\mathcal{O}(\epsilon)$, the off-diagonal element $\epsilon$ is written in terms of the covariance of $\bracket{ \xi_{g}\xi_{g+1} }$ as 
\begin{align}
%\epsilon = \eta / \sigma^{-2} = r^{-(g+1)} \sigma^{-2} \bracket{ \xi_{g}\xi_{g+1} }. \label{epsilon-rCov}
\epsilon = \sigma^{-2} \bracket{ \xi_{g}\xi_{g+1} }. \label{epsilon-rCov} %= \eta / \sigma^{-2} 
\end{align}
The covariance of $\xi_{g}$ is written in terms of the covariance of $J_{g}$ according to Eq.~(\ref{J-xi}) using Wick's theorem: 
\begin{align}
&\bracket{J_{i} J_{j}} - \bracket{J_{i}} \bracket{J_{j}} \nonumber\\
&= \mathrm{e}^{ \mu_{i} + \mu_{j} } \left( \bracket{\mathrm{e}^{ \xi_{i} + \xi_{j} }} - \bracket{\mathrm{e}^{\xi_{i}}} \bracket{\mathrm{e}^{\xi_{j}}} \right) \nonumber\\
&= \mathrm{e}^{ \mu_{i} + \mu_{j} }\left( \sum_{w=0}^{\infty} \frac{1}{w!} \bracket{ \xi_{i} + \xi_{j} }^{w} -  \sum_{u,v=0}^{\infty} \frac{1}{u! v!} \bracket{ \xi_{i} }^{u} \bracket{ \xi_{j} }^{v} \right) \nonumber\\
&= \mathrm{e}^{ \mu_{i} + \mu_{j} } \left( \sum_{l=1}^{\infty} \sum_{m,n=0}^{\infty} 
\frac{1}{(l+2m)!(l+2n)!} l! \frac{(l+2m)!}{2^{m} m! l!} \frac{(l+2n)!}{2^{n} n! l!} 
\bracket{\xi_{i}\xi_{j}}^{l} \bracket{\xi_{i}^{2}}^{m} \bracket{\xi_{i}^{2}}^{n}
\right) \nonumber\\
&= \mathrm{e}^{ \mu_{i} + \mu_{j} } \mathrm{e}^{ \frac{1}{2}(\sigma_{i}^{2} + \sigma_{j}^{2}) } (\mathrm{e}^{\bracket{ \xi_{i}\xi_{j} }} - 1) \nonumber\\
&= \bracket{J_{i}} \bracket{J_{j}} (\mathrm{e}^{\bracket{ \xi_{i}\xi_{j} }} - 1). 
\end{align}
Therefore, Eq.~(\ref{epsilon-rCov}) now reads 
\begin{align}
\epsilon = \sigma^{-2} \ln \frac{ \bracket{J_{g} J_{g+1}} }{ \bracket{J_{g}} \bracket{J_{g+1}} }. \label{epsilon-J}
\end{align}
Substituting Eq.~(\ref{epsilon-J}) into Eq.~(\ref{threshold-epsilon}), 
we have the shift of the threshold of the tipping point $\beta_{\mathrm{ex}}$ in the form  
\begin{align}
%\beta_{\mathrm{ex}} &= \overline{k}^{-1} \lim_{g \rightarrow \infty} \left( \frac{ \bracket{\beta_{g}} \bracket{\beta_{g+1}} }{ \bracket{\beta_{g} \beta_{g+1}} } \right).
\beta_{\mathrm{ex}} &= \overline{k}^{-1} \left( \frac{ \bracket{\beta_{g}} \bracket{\beta_{g+1}} }{ \bracket{\beta_{g} \beta_{g+1}} } \right) \nonumber\\
&= \overline{k}^{-1} \left[ 1 + \rho(\beta_{g}, \beta_{g+1}) \frac{ V(\beta_{g}) }{ \bracket{\beta_{g}}^{2} } \right]^{-1}, \label{threshold-beta} %\\
%&= \overline{k}^{-1} \left[ 1 + \rho(\beta_{g}, \beta_{g+1}) \frac{ V(\beta_{g}) }{ \bracket{\beta_{g}}^{2} } \right]^{r^{-2}}, \label{threshold-beta} %\\
\end{align}
where $\rho(\beta_{g}, \beta_{g+1})$ is the correlation coefficient which varies from $-1$ to $1$ and $V(\beta_{g})$ is the variance of $\beta_{g}$. 

\begin{figure}[t]
\begin{center}
\includegraphics[width=0.7 \textwidth]{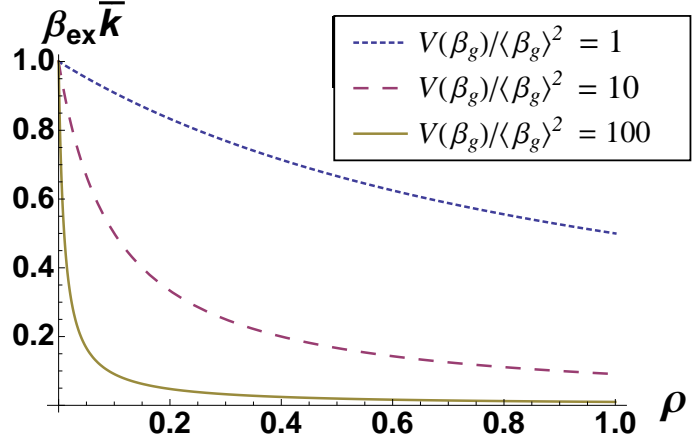}
\end{center}
\caption{
(Color online) 
The dependence of the tipping point of the viral spreading, Eq.~(\ref{threshold-beta}), as a function of the correlation coefficient $\rho(\beta_{g}, \beta_{g+1})$. The result of Eq.~(\ref{Independent2}) corresponds to the case where $\rho = 0$. %We set $\overline{k} = 1$. %and we ignored the factor $\mathcal{O}(\epsilon^{2})$ of $r^{2}$ at the exponent. 
}
\label{ThresholdPlot}
\end{figure}

%Ignoring the factor $\mathcal{O}(\epsilon^{2})$ of $r^{2}$ at the exponent, 
We exemplify the behavior of Eq.~(\ref{threshold-beta}) in Fig.~\ref{ThresholdPlot}. 
If $V(\beta_{g}) / \bracket{\beta_{g}}^{2} \sim \mathcal{O}(1)$, the tipping point would be lowered only up to a half of the case of the independent process, while it is lowered significantly in the case where $V(\beta_{g}) / \bracket{\beta_{g}}^{2} \gtrsim \mathcal{O}(10^{2})$; even when $\rho(\beta_{g}, \beta_{g+1}) = 0.2$, the spreading is about twenty times more likely to go viral than the uncorrelated case.

\section{Discussion and Conclusion}
When we discuss the viral spreading, the average of the spreading rate is not the only significant factor, but its fluctuation and the correlation may also play important roles. 
Equation (\ref{threshold-beta}) means that the tipping point where the spreading goes viral is shifted owing to the correlation $\rho(\beta_{g}, \beta_{g+1})$ of the spreading rates between the generations. 
The larger the variance $V(\beta_{g})$ of the spreading rate is compared to the square of its average $\bracket{\beta_{g}}$, the easier it is to make the spreading go viral. 
On the other hand, it is hopeless to expect the information spreading with very narrow variance of the spreading rate to go viral, 
unless it is constantly very close to the tipping point of the uncorrelated case, $\beta_{\mathrm{ex}} = \overline{k}^{-1}$.

We defined the tipping point of the viral spreading as a theoretical guideline of the information spreading on an online social network such that the information reaches the nodes which are extremely far from the seed node. 
We showed how the correlation between the nodes enhance the chance of the viral spreading. 
Although we used a perturbation expansion with respect to the off-diagonal matrix element $\epsilon$ in Eq.~(\ref{epsilon-expansion}), 
its higher-order expansion is straightforward. 
Note that $\epsilon$ cannot be too large, in other words, $\rho(\beta_{g}, \beta_{g+1})$ cannot be close to one,
 in order to retain the positivity of the covariance matrix $\Sigma$, 
which also validates the perturbation expansion. 
We numerically showed that the true tipping point may be even lower than the current result of the perturbative approach.

For Twitter, the tipping point would be unrealistically far to reach without the correlation between the generations. 
The significant change of the tipping point due to the correlation seems to be essential in understanding the mechanism 
why such postings sometimes diffuse extremely far from the seed user.

The tipping point (\ref{threshold-beta}) may be still far to reach even after taking into account the correlation effect. 
The assumptions which we made on the underlying network such as the homogeneity of distribution and the infinite path length may cause the change of the estimation of the tipping point. 
In order to analyze the spreading more precisely, removing these assumptions is an interesting future problem. 
The inhomogeneity of the underlying network such as the community structure may restrict information spreading within a community,  while the assumption of infinite path length itself should not be essential for the evaluation of the tipping point, because 
the spreading rate is basically a local quantity of a generation (although it may be correlated with the neighboring generations). 
Note that, even though the average path lengths are usually very short for many networks in real world \cite{Dorogovtsev}, 
the path length of the spreading can be much longer than the average path length of the underlying network, 
because the spreadings do not always occur along the shortest paths \cite{Liben-Nowell08,Golub10,Bakshy11,Raj}.

It may also be important to consider the dependence of the spreading rate on the generation; for example, decrease of the average spreading rate and the strength of correlation will raise the tipping point or restrict any spreading to a finite size. 
Although the correlation decreases along the generations in the present analysis as shown in Eq.~(\ref{CovMat}), 
in reality, the covariance matrix is not necessarily of the form which we considered here.

While a few improvements on the effects by the network structure is possible, 
considering temporal effects on the dynamics may also be an interesting issue. 
Since many results have been revealed on temporal networks recently \cite{Iribarren11,Vazquez07,Karsai11,Holme12}, 
inclusion of such features may lead us to a deeper understanding of the information spreading in online social networks. 
When a precise dataset is available, a quantitative analysis with a realistic model which takes account of the above points will be fruitful.

The range of application of the above analysis may be broader than the information spreading in online social networks. 
Although it is not easy to guess what kind of spreading has a non-trivial probability distribution for the spreading rate from current knowledge, 
as a close example, we can think of the propagation of a phishing attack due to an electric mail with a virus which sends copies of the mail to all the people in an infected user's address book.

\section*{Acknowledgements}
This work was partially supported by the Aihara Innovative Mathematical Modelling Project, the FIRST program
from JSPS, initiated by CSTP, Global COE Program ``the Physical Sciences Frontier", MEXT, Japan, and Kawarabayashi Large Graph Project, ERATO, JST. 
%Even though the assumption for the underlying network that the graph has a loopless structure with a homogeneous distribution and has infinite path length may be too strong to apply the result of the present work to the real social networks, the significant drop of the tipping point of the tweet spreading seems to be essential to understand the mechanism why such postings sometimes diffuse explosively. 



%% References


\bibliographystyle{elsarticle-num}
\bibliography{<your-bib-database>}



\end{document}